\documentstyle[12pt,aaspp4]{article}

\newcommand {\hii}{\ion{H}{2}}

\newcommand {\ha}{H$\alpha$}
\newcommand {\oiii}{[\ion{O}{3}]}
\newcommand {\sii}{[\ion{S}{2}]}
\newcommand {\degr}{\arcdeg}
\newcommand {\amin}{\arcmin}
\newcommand {\asec}{\arcsec}

\def \ampt      {\farcm}
\def \aspt      {\farcs}

\newcommand {\et}{{et al.}}
\newcommand {\eg}{{e.g.,}}

\newcommand {\rosat}{{\it ROSAT\ }}
\newcommand {\einst}{{\it Einstein\ }}

\tighten

\begin{document}

\title {Supernova Remnants in the Magellanic Clouds III: \\
An X-ray Atlas of LMC Supernova Remnants}
\author{Rosa Murphy Williams, You-Hua Chu, and John R. Dickel}
\affil{Astronomy Department, University of Illinois at Urbana-Champaign,
Urbana, IL 61801}
\author{Robert Petre}
\affil{NASA/GSFC code 666, Greenbelt, MD 20771}
\author{R. Chris Smith}
\affil{Cerro Tololo Inter-American Observatory, La Serena, Chile}
\author{Maritza Tavarez}
\affil{Astronomy Department, University of Michigan, Ann Arbor, MI 48109-1090}

\begin{abstract}
We have used archival \rosat\ data to present X-ray images of thirty-one
supernova remnants (SNRs) in the Large Magellanic Cloud (LMC).  We have 
classified these remnants according to their X-ray morphologies, into the 
categories of Shell-Type, Diffuse Face, Centrally Brightened, Point-Source 
Dominated, and Irregular.  We suggest possible causes of the X-ray emission 
for each category, and for individual features of some of the SNRs.
\end{abstract}


\section{Introduction}

One of the most influential factors in the shaping of the interstellar 
medium (ISM) is the input of energy from supernova explosions and the 
subsequent development of their remnants.  The long-term evolution of a 
supernova remnant (SNR) is subject to both the details of the original 
explosion (initial energy input, explosion asymmetries, etc.) and the 
characteristics of its surrounding environment.  A unique opportunity 
to study SNRs from different initial conditions, in a wide variety of 
environments, is provided by the Large Magellanic Cloud (LMC). The LMC, 
at a known distance of $\sim$ 50 kpc\ \markcite{F91}(Feast 1991), 
allows us to circumvent Galactic uncertainties in distances and 
absorptions, while retaining reasonable spatial resolution. Taking
advantage of these properties of the LMC, we have begun a systematic
study of SNRs in the Magellanic Clouds. In previous work, we have made 
multiwavelength studies of the colliding remnants DEM L 316 \markcite{W+97}
(Williams \et\ 1997; Paper I) and for two SNRs, N 11 L and N 86, which 
show breakout structures \markcite{W+98}(Williams \et\ 1998, Paper II).

The emission of SNRs at X-ray wavelengths traces both the hot gas 
immediately behind the shock front (including the reverse shock, which
predominates in younger SNRs), and in some cases, that in the central 
cavity. For instance, cloudlet evaporation in an SNR interior can raise 
the density of the hot gas, thus raising the X-ray surface brightness of 
the hot interior to observable levels.  This phenomenon is an example 
which shows that the X-ray morphology can be a useful tool for diagnosing 
the physical properties of an SNR.  The extensive database provided by 
the archive of \rosat\ observations gives us the opportunity to study 
the X-ray emission of the LMC SNRs as a group. A variety of X-ray 
morphologies are observed in LMC SNRs.  In this paper (Paper III), we 
present an X-ray atlas of those SNRs and discuss their respective 
morphologies.  Our next paper (Paper IV) will discuss the X-ray spectra 
of these remnants obtained from the \rosat\ PSPC.

In \S2 of this paper, we will describe the X-ray dataset.  In \S3, we will 
define categories of SNRs according to  their X-ray morphology, and classify 
the LMC SNRs in terms of these categories.  \S4 will discuss the implications 
of the different morphologies, and \S5 will summarize our findings.

\section{X-ray Dataset}

Our X-ray dataset is taken from the \rosat\ archive, and includes all 
known LMC SNRs for which data were available.  The list of known 
SNRs was based on the papers of Mathewson \et \markcite{M+83}
\markcite{M+84}\markcite{M+85} (1983, 1984, 1985) with additions for 
newly discovered SNRs as per Chu \et \markcite{C+93}\markcite{C+95}
\markcite{C+97}(1993, 1995, 1997) and Smith \et\ \markcite{S+94}(1994), 
as well as the recent SN 1987A.  Only those SNRs with \rosat\ data 
available are listed in Tables 1, 2, and 3, whereas for completeness, 
Table 4 contains entries for the full list of known LMC SNRs, based 
on the references above.

For most of the objects discussed in this paper, we used data from the
\rosat\ High Resolution Imager (HRI).  The HRI is  sensitive in the energy 
range of 0.1 $-$ 2.0 keV, with a field of view of 38\amin $\times$ 38\amin. 
The in-flight angular resolution of the HRI and the X-ray telescope system 
is approximately 5\asec $-$ 6\asec\ (half power diameter; \markcite{R91}
\rosat\ Mission Description 1991).  To increase the signal-to-noise ratio 
(S/N) and to preserve the angular resolution, we binned the images to 
1\asec\ pixels, then smoothed them with a Gaussian of $\sigma$=2 pixels, 
for an effective resolution of approximately 7\asec.  Images for which the 
resulting surface brightness was too low for ready classification were 
re-binned to lower resolutions until the counts per pixel were sufficient 
to distinguish the features of the object.  

Four objects, N 86, N 120, SNR 0454$-$672, and SNR 0453$-$669, had insufficient 
exposures with the HRI, but did have images available from the Position 
Sensitive Proportional Counter (PSPC).  The PSPC has an on-axis angular 
resolution of $\sim$~30\asec\ and a circular field of view 2\degr\ in 
diameter \markcite{P+87}\markcite{R91} (Pfeffermann \et\ 1987; \rosat\ 
Mission Description 1991).  The PSPC images were binned to 5\asec\ pixels 
and smoothed with a Gaussian of $\sigma$=2 pixels; the resulting image 
resolutions are $\sim$ 38\asec.

The images were processed and analyzed using the IRAF/PROS package.  
Table 1 lists coordinates for the SNRs, and the instrument, \rosat\ 
sequence number, and exposure time for each image used in this atlas.  
Table 2 lists the background-subtracted total count rate, average 
brightness, and estimated X-ray size of the SNR. The count rate was 
calculated by determining the average counts per second within the SNR 
itself (as identified from its X-ray morphology) minus the background 
contribution estimated from a similar off-source region; the average 
brightness is this count rate divided by the estimated area of the remnant 
on the sky.  Telescope vignetting is not considered to be a significant
factor; most of the observations used here were on-axis, and, at 1 keV, 
vignetting is less than 10\% for  objects up to 15\amin off-axis.
Likewise, at 1 keV, mirror scattering causes about 6\% of incident
photons to fall outside a radius of 10\asec; this effect is not expected
to significantly affect the count rates or estimated resolutions given
in this paper \markcite{R97}(\rosat\ High Resolution Imager Calibration 
Report 1997, pp. 7, 22).

The count rate can be converted to an energy flux for the 
0.1$-$2.4 keV range, using an Energy Conversion Factor (ECF) based on the 
parameters of the object spectrum \markcite{R91}(\rosat\ Mission Description 
1991).  This flux can be converted to a luminosity, assuming a distance of 
50 kpc to the remnants.  For a Raymond-Smith plasma with a temperature of 
$\sim$0.5 keV and an absorption column density of $\sim$10$^{21}$ cm$^{-2}$, 
the ECF is 0.15 $\times$ 10$^{11}$ counts cm$^{2}$ erg$^{-1}$. Thus, the 
count rate could be multiplied by a factor of 2$\times$10$^{37}$ to convert 
it to an X-ray luminosity in erg s$^{-1}$.

Figures 1 and 2 show the complete atlas of those LMC SNRs for which \rosat\
observations were available.  Figure 1 contains images for which the X-ray
data were sufficient for morphological classification; Figure 2 shows the
remainder.  Contours in these images are set as multiples of 2$\sigma$\ 
above the background, beginning at 2$\sigma$\ and doubling thereafter.  
In those cases where the emission is particularly strong, these figures 
are multiplied by ten, so that the contour levels begin at 20$\sigma$\ 
above the background and double thereafter.  In cases such as N 158A, which 
contain both bright pointlike and faint diffuse emission, both systems are 
used.   For comparison, contour maps of a ``clean" unresolved source are 
shown in the \rosat\ High Resolution Imager Calibration Report \markcite{R97}
(1997, p.17). The effective resolution of the data is plotted as a black
dot in the lower right of each image; this effective resolution is the
on-axis half power diameter convolved with the Gaussian smoothing of the 
image.  Table 3 lists relevant information for each image, such as the 
pixel size, image resolution, brightness at the lowest contour level (in 
total counts per pixel), the $\sigma$\ value for each image, and the 
contours in $\sigma$\ multiples.

\section{X-ray Classification of SNRs}

The X-ray morphology of an SNR can provide a considerable amount of
information about that SNR. A point (unresolved) source may indicate
the presence of a neutron star. A bright limb suggests some combination 
of a fast shock and a moderately dense ambient medium.  Bright X-ray 
patches may indicate that the shock is encountering ISM clumps, or 
may result from the later evaporation of such shocked clumps in the 
hot interior of the remnant.  SNRs in which such clumps are sufficient 
to dominate the emission may even develop a centrally brightened X-ray
morphology.  Our goals in this paper are to present the X-ray 
morphologies of the LMC SNRs, classify them according to distinguishing
characteristics, and attempt to explain these categories in terms of 
possible physical causes.

Table 4 lists our classifications for the LMC remnants.  Also listed 
are those SNRs for which \rosat\ data were unavailable, or were 
insufficient for classification.  We have used the following primary 
X-ray characteristics in our classification:

\begin{list}{}{\itemsep 0pt \itemindent-15pt}
\item Shell: limb is near-complete and is brighter than face
\item Diffuse Face: face as bright as shell; limb sometimes indistinct 
	from face
\item Centrally Brightened: center of face notably brighter than limb
\item Peaked Emission: emission dominated by central bright source of 
	small diameter with respect to that of the entire SNR
\item Irregular: incomplete or nonexistent shell, patchy emission
\end{list}

In addition, we wish to include designations for secondary X-ray features
within a remnant. These features are defined as follows:

\begin{list}{}{\itemsep 0pt \itemindent-15pt}
\item Gap: region of weaker emission in near-complete shell
\item Hotspot: proportionally small region of bright diffuse emission
\item Bright Limb: elongated bright region along shell
\item Non-Limb Feature: patch of emission away from the limb
\item Elongated Extension: protrusion beyond shell limb
\end{list}

Several earlier papers have used the substantial \einst\ database to 
classify the X-ray morphologies of SNRs.  Mathewson \et\ \markcite{M+83}
(1983) identified three categories: a ``limb-brightened shell," and a 
``uniform... sometimes patchy" distribution, which are very similar
to our own Shell and Diffuse Face categories, and a ``centrally peaked"
morphology which the authors identified as corresponding to Crab-type 
X-ray emission. In general we equate this last category with 
our Peaked Emission classification. It should be noted that the SNR 
N 103B is listed as centrally peaked by Mathewson \et, whereas 
improved data place this remnant in the Shell category.

Seward \markcite{S90}(1990) likewise categorized Galactic remnants according 
to their X-ray morphologies as seen in \einst\ images; these categories 
included ``shell," ``filled center, plerionic," and ``irregular," with 
additional notes for unresolved central objects and known neutron stars. 
For the most part, these categories overlap with our Shell, Peaked Emission 
or Centrally Brightened, and Irregular classifications.  Remnants showing 
the sort of patchy emission which we classify as Diffuse Face are classified 
by Seward as a combination of ``shell" and ``filled center" or ``central 
(unresolved) object;" some are listed as``composite" remnants.

\subsection{Shell -- Type SNRs (Fig. 1a-f)}

The SNRs showing largely shell-type emission include DEM L 71, SNR 0509$-$675, 
N 103B, SNR 0519$-$690, N 132D, and  DEM L 316 B. Two features of this group 
are particularly notable. The first point is that with the exception of 
DEM L 316 B, this group is comprised of most of the younger LMC remnants.
Three of them, DEM L 71, SNR 0509$-$675, and SNR 0519$-$690, have Balmer-line
dominated optical emission. Such emission typifies remnants of Type Ia SN,
in which a collisionless shock moves into partially neutral ISM 
\markcite{CR78}(Chevalier \& Raymond 1978), and therefore we do not expect 
the forbidden-line radiation which is associated with the radiative cooling 
region behind the SNR shock \markcite{RS77}(\eg\ Raymond \& Smith 1977).

A second point of interest is that in all of these cases, the  X-ray 
emission is not symmetric around the shell; SNR 0509$-$675 and N 103B show 
bright limb hotspots; SNR 0519$-$690 and  DEM L 316 B have gaps breaking 
the symmetry of their shells; N 132D shows not only such a gap, but also 
X-ray extensions pointing away from the gap itself.  DEM L 71 has perhaps
the greatest deviation from the simple shell morphology; the limb features 
two bright arcs opposite one another, and a hotspot near the center of the 
face of the remnant. 

\subsection{Diffuse Face SNRs (Fig. 1f-n)}

There are a subset of LMC SNRs that neither fit the simple shell-type 
category nor meet the criteria for the centrally brightened SNRs.  We
label these intermediate cases as ``diffuse face" SNRs. The distinguishing
characteristic for this category is the presence of significant X-ray 
emission over the face of the remnant, without clear limb-brightening. 
Objects in this category include the remnants SNR 0453$-$685, SNR 0520$-$694, 
N 49B, N 49, SNR 0534$-$699, N 63A, DEM L 316 A, and SNR 0548$-$704. 

There appear, in addition, to be subgroups in this category. SNRs 0453$-$685, 
0520$-$694, 0534$-$699, and 0548$-$704 each have a relatively even distribution 
of surface brightness over the remnant. In contrast, the remnants N 49B and 
N 63A are significantly brighter to one side and have brighter hotspots
along the rim. The latter morphology may in fact arise from an encounter
of one side of a shell-type remnant with denser ISM; the encounter surface,
at an angle to the line of sight, appears as a brighter patch on one side
of the remnant.  DEM L 316 A appears to be one of these asymmetrically 
limb-brightened remnants, possibly due to an encounter with the neighboring
remnant DEM L 316 B.

\subsection{Centrally Brightened SNRs (Fig. 1o-q)}

To date, the only clear example we have found of an LMC SNR that shows 
centrally brightened X-ray emission is N 206. 
While the remnants N 120 and 0454$-$672 also seem to have central 
concentrations of X-ray emission, these SNRs are too faint for unambiguous 
morphological classification.  It should also be noted that many of the 
smaller remnants, which have particularly bright hotspots along the limb, 
might be mistaken for centrally brightened SNRs on casual inspection. 
The distinguishing characteristics of the centrally brightened SNRs, 
however, are that the peak of the X-ray emission is located away from 
the outer shell, as seen in both X-ray and optical emission; and that 
the X-ray peak does not have corresponding emission at radio and optical 
wavelengths.  When taken in the context of other wavelengths, these are
commonly called ``mixed morphology" SNRs (\eg\ Rho \& Petre 1998).

\subsection{Peaked-Emission SNRs (Fig. 1r-s)}

The X-ray emission from SNRs N 157B and N 158A appears to be dominated by
nonthermal emission from a central small-diameter source.  Both of these 
SNRs contain known pulsars (\eg\ \markcite{M+98}Marshall \et\ 1998; 
\markcite{WG98} Wang \& Gotthelf 1998; \markcite{S+84} 
Seward, Harnden \& Helfand 1984).  As the X-ray spectra of both are
predominantly nonthermal (power-law), it is reasonable to suggest that
the bulk of the X-ray emission is powered by the central neutron star.

However, there does appear to be additional emission surrounding the 
small-diameter source in each SNR. Some of this emission may be 
attributable to synchrotron nebula surrounding the central pulsar, but 
some appears to be diffuse emission, possibly arising from the expanding 
shock front of the SNR.  N 158A is almost certainly such a ``composite"
remnant, with a clear shell structure around the central peak; N 157B
appears to have less well-defined outer structure, and may be in 
transition to a composite morphology.

\subsection{Irregular SNRs (Fig. 1t-x)}

As always, some objects fall outside the standard categorizations.
These may have unique features due to specific local conditions, or
may simply be the result having reached an age when the X-ray emission
is no longer sufficient to show any sort of pattern.  Objects in this
category include the SNRs N 23, DEM L 238, DEM L 299, the Honeycomb 
Nebula, N 11L, and N 86.  The elongated X-ray structure of N 23, which 
follows part of the optical limb, may be an extreme case of the asymmetric 
limb-brightening suggested for N 49B and N 63A.   DEM L 238 and DEM L 299 
appear to have rough shell structures somewhat within the optical shells, 
and may be similar to DEM L 316 B in representing a backfilling of shocked 
material. The Honeycomb Nebula is irregular in optical structure as well; 
it is unsurprising that the X-ray emission is similarly irregular.  The 
X-ray emission of N 11L is thought to follow a breakout of hot gas from this 
remnant \markcite{W+98}(Paper II), resulting in the unusually 
extended morphology.  Similarly, N 86 is thought to show one or more 
breakouts, creating an extended multi-lobed structure \markcite{W+98}
(Paper III).

\section{Physical Implications}

\subsection{Schematic Evolution of X-ray Morphology}

The X-ray emission in an SNR can be generated by several different 
mechanisms.  The bulk of the X-rays from most SNRs appears to consist of 
thermal line emission, with an additional contribution from thermal 
bremsstrahlung emission from radiative cooling of hot gas behind the 
shock (Ellison \et\ 1994, and references therein).   Nonthermal X-rays 
may also be produced by synchrotron radiation at the shock front, as the 
shock compresses magnetic fields and accelerates relativistic electrons 
(Reynolds 1988).  If a pulsar is present, it may inject relativistic 
particles into its surroundings via a ``pulsar wind",  to form a 
``pulsar-powered nebula" which radiates synchrotron emission (\eg\ 
Kennel \& Coroniti 1984).

In a very young remnant, in which the supernova ejecta still dominates,
the bulk of the X-ray emission arises from the reverse shock traveling
back through this ejecta.  Later in the young remnant's development, the 
reverse shock ceases to dominate the emission; then the fast expanding 
shock will have the brightest X-rays along the outer rim, as the front 
shocks the ambient, relatively dense ISM to high temperatures (\eg\ 
Raymond 1984 and references therein).  In remnants with more powerful 
shocks, nonthermal X-rays will also contribute to the emission along the 
shock front (Reynolds 1988).  Material within the hot interior of such 
SNRs will be too rarefied to have sufficient surface brightness for the 
easy detection of X-rays. These lead to the expected ``shell-type" X-ray
morphology.   

As the outer shock proceeds, cloudlets in the ambient medium may be left 
behind when the shock passes them. These cloudlets may subsequently evaporate 
into the hot interior, producing detectable X-ray emission near these local 
maxima in the interior density.  For example, N 63A shows a bright spot
in X-rays corresponding to cloudlets behind the shock seen in optical
emission \markcite{C+98}(Chu \et\ 1998). Such cloudlets may produce the 
sort of morphology here called ``Diffuse Face" SNRs.  

When interior cloudlets are particularly prevalent, or the outer shock less
prominent, this process may cause the  X-ray emission to peak toward the 
center of the remnant, as the hot interior becomes dense enough for its 
emission to overwhelm that from the shock front itself.  This may lead to 
the ``centrally brightened" morphology, as modeled by White \& Long 
\markcite{WL91}(1991). However, it is possible that an intermediate form of 
such brightening can arise.  This could range from a ``thick shell," where 
X-ray emission comes from a wide band of post-shocked material, to an even 
distribution of emission over the remnant, to the central peaking of X-ray 
emission.  A cartoon of these schematic stages of development is shown in 
Figure 3.

On a cautionary note, it should be pointed out that the ``mixed morphology" 
remnant structure, which includes Centrally Brightened X-ray emission, can 
be attributed to other processes than the  White \& Long \markcite{WL91}
(1991) model.  A competing model, summarized in Rho \& Petre \markcite{RP98}
(1998, and references therein) suggests that the central emission is 
``fossil" radiation remaining from the SN explosion.  In this model, the 
central emission is bright enough to be observable due to a high ambient 
density.  At young ages, the central emission is overwhelmed by that from 
the outer shock, but as the remnant enters the radiative stage, the 
interior emission comes to dominate.  This model does not include clumpy 
emission interior to the SNR, however; the Diffuse Face morphology cannot 
be explained through this mechanism, although other external factors 
could produce such morphologies.

\subsection{Observed X-ray Morphologies of LMC SNRs}

We can compare the schematic stages of remnant development 
described above to actual SNR morphologies, as seen at X-ray wavelengths.
We can include additional data from the X-ray spectra of some objects for
a more complete picture.

The younger, smaller remnants, such as SNR 0509$-$675, SNR 0519$-$690, and
N 103B, tend to show a clear shell-type morphology in X-rays. Some of this
X-ray emission does indeed appear to be nonthermal, as shown in a recent 
study by Petre \et\ \markcite{P+97}(1997), who found a greater tendency to 
have hard X-ray emission in small-diameter (presumably, younger) SNRs than 
in larger ones.  For some of these remnants, the hard X-ray emission appears 
to be predominantly nonthermal, as is reasonable for young SNRs with fast
shock fronts.

The Diffuse Face category covers a wide variety of X-ray morphologies.
Some remnants appear to have limb-brightened emission that extends over 
the face of the remnant; N 49 and N 63A are examples of this sub-category.  
Others, such as N 49B and SNR 0453$-$685, have patchy emission over the 
entire face. These variations can plausibly be explained as a range
of transitional stages, ranging from the extensions from the shell seen
in N 49 and N 63A, to the more ``uniform" distribution in (presumably)
older remnants.

We would expect to find the Centrally Brightened SNRs at the far end of
this range, as the clump emission raises the central surface brightness
of the remnant to the point where the interior dominates the X-ray emission.  
N 206 is perhaps the best example of the Central Brightening in X-rays, 
although SNR 0454$-$672 and N 120 are also candidates.  Note that it is 
important to consider the morphologies at other wavelengths with regards 
to this category, as the Mathewson \et\ \markcite{M+83}(1983) classification
of N 103B shows.  From the X-ray morphology alone, especially if the X-ray 
spatial resolution and sensitivity are poor, one may mistake an SNR with a 
bright limb or a breakout for a Centrally Brightened remnant.

A loose correlation between remnant size and some of the morphological 
classes is shown in Figure 4.  The smallest SNRs are all Shell-type, and 
the Shell-type SNRs are all at the lower end of the range of diameters. 
Conversely, the remnants known or suspected to be Centrally Filled are all 
at the upper end of the range of X-ray sizes.  Unsurprisingly, the Diffuse 
Face remnants fill most of the range, overlapping with both Shell-type and 
Centrally Filled, but concentrated between the two extremes.  Those Diffuse 
Face SNRs still appearing to be somewhat shell-like fall neatly between the 
majority of Shell-type and the bulk of other Diffuse Face SNRs.  While this 
is of course not conclusive, and suffers from small-number statistics, it 
indicates that the morphologies do reflect an evolutionary trend.  This is
strengthened by the fact that the three smallest, Shell-type SNRs are young
enough to be ejecta-dominated \markcite{H+95}(Hughes \et\ 1995). The ages 
calculated by Hughes \et\ \markcite{H+98}(1998) through X-ray spectral 
fits are less helpful for comparison, as they describe objects all in the 
region where the Shell-type and Diffuse Face SNRs overlap in size, but do 
confirm the comparative youth of these relatively small-diameter SNRs.

Two categories do not fit well within the schematic of SNR development
given above. The Point Source Dominated SNRs have X-ray emission primarily
arising from synchrotron radiation powered
by a central pulsar, making the X-ray morphology 
elsewhere in the remnant difficult to distinguish.  The fainter, diffuse 
X-ray emission from N 158A does suggest a shell, while that of N 157B is 
Crab-like.  Pulsar-dominated remnants tend to be spectrally dominated, to 
a large degree, by nonthermal X-ray emission from the central small-diameter
source, making their emission fairly distinctive.

The Irregular SNRs may have morphologies more determined by the details of
the surrounding ISM than by their stage of development; the Honeycomb SNR,
N 11L, and N 86 are all examples of this. Others such as DEM L 238 and 
DEM L 299 may be evolved to the point where most of the emission is due
to backfilled clumps rather than recent shocks, while still others may 
simply have too low surface brightness to distinguish any but the brightest 
patches of the X-ray emission.

One problem with morphological classification is the possibility of 
projection effects.  Remnants which are encountering spots of higher 
density material in the direction toward our line of sight may have areas 
along the face which are brightened in X-rays. These patches of emission 
may create a morphology which would be easily confused with the ``Diffuse 
Face" class of SNRs.  It is also possible that a particularly bright spot 
on the limb may, due to the shell curvature or poor resolution, appear to 
brighten part of the face as well, as with N 103B and, possibly, N 49.

\subsection{Multiwavelength Morphological Comparison of LMC SNRs}

To properly describe the structure of these SNRs, we compare the X-ray
morphology to the morphologies at other wavelengths.  We chose examples
from each X-ray morphological category, to examine in comparison with 
its radio and optical emission.  These examples were chosen according to 
the availability of good radio and optical data; in the case of the 
``Centrally Brightened" category, as only N 206 definitely fits that 
class, it was chosen as the example object.  Optical (\ha) images used for
this comparison were taken using the 1.5 meter telescope at the Cerro 
Tololo Inter-American Observatory (CTIO; \markcite{S98}Smith 1998).  The 
radio images were obtained with the Compact Array of the Australia 
Telescope (ATCA; \markcite{DM98}Dickel \& Milne 1998).

As an example of this multiwavelength comparison for a shell-type remnant, 
we consider the SNR N 103B.  The X-ray brightening seen on the western limb 
of this remnant corresponds to a set of bright \ha\ knots (Fig. 5a). While 
this western side of the shell is fairly well defined in \ha, the rest of 
the shell is optically faint.  The same X-ray bright portion of the limb 
corresponds to a bright region in the radio image (Fig. 6a), which shows 
a complete limb-brightened shell with significantly brighter emission on 
the western side.

We use both N 49 and N 49B as comparison objects for the Diffuse Face 
category, in order to illustrate the broad range of such objects.
In \ha, N 49 appears as a tangle of internal filaments surrounded by
fainter diffuse emission.  The brightness of these filaments roughly
corresponds to the brighter regions in X-rays. The radio emission is
more shell-like, without clear correspondence with the optical filaments
in most cases.  It should be noted that the northern spot of emission in 
N 49, thought to correspond to the soft gamma-ray repeater (SGR) 0525$-$66  
\markcite{M+96}(\eg\ Marsden \et\ 1996), has no radio or optical counterpart. 
A second bright spot, however, on the southeastern limb, does in fact 
correspond well to a bright spot in the radio (Fig. 6b) and nearby 
limb-brightening in both radio and \ha\ (Fig. 5b, 6b).

The \ha\ image of N 49B (Fig. 5c) shows little or no shell structure, 
having instead several patches of bright optical emission and fainter 
diffuse emission overall.  The brightest patch of X-ray emission, on the 
southern limb, corresponds to one of these bright optical regions, but 
over the rest of the remnant the X-ray and \ha\ emission appear to be 
uncorrelated.  An ATCA image for N 49B (Fig. 6c) does show a rough 
shell structure, with diffuse emission over the interior.  Two bright
patches of X-ray emission, on the northern limb and on the eastern side
of the face, are matched by similar patches of radio emission; however,
the X-ray spot on the southern limb shows no associated radio emission.

N 206 was chosen to illustrate the X-ray morphological category of 
Centrally Brightened SNRs.  The \ha\ image for N 206 (Fig. 5d) shows a 
clear limb-brightened shell, while the X-ray emission peaks toward the 
center of the remnant. This fulfils the criteria for a ``mixed-morphology" 
SNR, as described in \S 3.3 above. No ATCA images are available for any 
of the Centrally Brightened X-ray candidates, although ATCA data for 
N 206 are currently being analyzed. 

We use N 157B as an example of centrally peaked emission. In \ha\
(Fig. 5e), N 157B is not remarkable; a small optical knot corresponds to the 
position of the X-ray point source, but the remnant is otherwise difficult 
to distinguish from the \ha\ emission of the surrounding \hii\ region
\markcite{C+92}(\eg\ Chu \et\ 1992).  
At X-ray and radio (Fig. 6d) wavelengths, however, the emission is dominated 
by a strong small-diameter source, corresponding to the presence of a pulsar 
within this SNR. Surrounding this central point is additional radio and X-ray 
emission, which forms an elongated structure largely to the northwest of 
the point source. 

As an example of an Irregular SNR, we used the Honeycomb Nebula, notable for 
its elongated and pocketed structure. The \ha\ (Fig. 5f), radio (Fig. 6e) 
and X-ray images are very similar; all three show a  ``ridge" of emission 
surrounded by fainter emission on all sides. A closer examination of the 
``ridge" shows it to be made up of numerous pockets of emission in all 
wavelength regimes. In the \ha\ images, these pockets are resolved into a 
``honeycomb" of small circular rings; the radio image is less clear, but 
is consistent with this picture.

Specific details in the X-ray morphology may, in some cases, be correlated 
with the characteristics of the surrounding medium. The bright limbs of 
objects such as N 23 and N 49 appear to coincide with the presence of 
higher-density material.  Several irregular remnants are strongly affected 
by the details of the surrounding ISM.  For example, the breakout structure 
seen in N 11L corresponds to an area with low optical emission, thought to 
represent a cavity in the shell of the  N 11 \hii\ region \markcite{W+98}
(Paper II). Another example is the Honeycomb Nebula, in which the 
``visible" portion is due to the encounter of the shock front with a porous 
sheet, creating the honeycombed structure \markcite{C+95}(Chu \et\ 1995).

\section{Summary}

The thirty-one remnants for which \rosat\ X-ray images could be obtained
are shown in Figures 1 and 2.  We have classified these remnants according
to their X-ray morphologies, into the categories of Shell-Type, Diffuse 
Face, Centrally Brightened, Point-Source Dominated, and Irregular.  We 
suggest that the observed X-ray emission from diffuse SNRs is initially 
generated at the shock front, where the freshly shocked million-degree 
gas is prominent; this results in a Shell-Type morphology.  As the front 
encounters and envelops clumps in the surrounding ISM, these clumps are 
shock-heated, and lag behind the SNR shock as the front continues to 
expand. Such clumps may evaporate into the interior of the SNR, causing 
local increases in the hot gas density; this increased density may in 
turn be sufficient to raise the surface brightness of interior X-rays to 
detectable levels.  This can result in the more distributed emission of 
the Diffuse Face morphology. As the interior density rises due to clump 
evaporation, emission from the high-temperature central interior may 
become most prominent, leading to a Centrally Brightened X-ray morphology.

Substantial deviations from this schematic concept of the development of
SNR X-ray morphology may occur for a number of reasons.  When a pulsar is
present, emission from this compact object may overwhelm that from elsewhere
in the remnant, creating a  Point-Source Dominated morphology.  In other
cases, variations in the ISM local to the remnant can have a substantial 
effect on the remnant's development.  This may in some cases result in 
completely Irregular morphologies, while in others it may lead to
specific features such as brightened regions, extensions from the shell,
or gaps in the shell continuity.  We note that such details may create
uncertainties in morphological classification, as may projection effects.
The comparison of the X-ray morphology with morphologies at other 
wavelengths can provide valuable insight as to the actual processes 
which have produced these morphologies.

\acknowledgements
This research has made use of data obtained through the High Energy 
Astrophysics Science Archive Research Center Online Service, provided by 
the NASA/Goddard Space Flight Center. We acknowledge the support of the 
NASA ADP grant NAG 5-7003.

\appendix
\section{Appendix: Individual SNR Properties and References}

{\it SNR 0450$-$709}\ (No \rosat\ data available). This object is one of 
the largest known SNRs in existence (\eg\ \markcite{J+98} Jones \et\ 1998), 
based on its optically-derived size of 104 $\times$ 75 pc \markcite{M+85} 
(Mathewson \et\ 1985).

{\it SNR 0453$-$685}\ (Fig.\,1i). This SNR shows patchy emission over the 
face of the remnant, with the brightest concentration located near the 
center.  X-ray spectroscopy of this SNR is provided by \markcite{H+98}
Hughes \et\ (1998).

{\it SNR 0453$-$669}\ (Fig.\,1x). This SNR was discovered by Smith \et\ 
\markcite{S+94}(1994) through \rosat\ observations of LMC \hii\ regions, 
confirmed by the high  \sii /\ha\ ratio in the region.  Velocity data
from Chu \markcite{C97}(1997) show it to have an expanding shell that 
is not spherically symmetric.  Due to a lack of HRI coverage of this 
object, we have used the \rosat\ PSPC image in this atlas. Its
asymmetric morphology, as well as the velocity data, suggest it may
be encountering dense material to one side.

{\it N 11L}\ (Fig.\,1u). This remnant is most notable for an apparent
breakout structure to the north.  In \ha\ and \sii, N 11L has a circular 
ring of projecting loop-like filaments to the northeast, which may be the 
result of a breakout into a local inhomogeneity (\eg\ \markcite{M87} 
Meaburn 1987).  Radio observations of N11L at 408 MHz by Mathewson and 
Clarke \markcite{MC73} (1973) and in  Molonglo Observatory Synthesis 
Telescope (MOST)  observations \markcite{MT+84} (Mills \et\ 1984) show 
a northern extension to N11L beyond the optical shell.  More recent
radio, X-ray and optical observations \markcite{W+98} (Paper II)
strengthen the evidence for a breakout in this remnant.

{\it SNR 0454$-$672}\ (Fig.\,1q). This SNR was also discovered by Smith \et\ 
\markcite{S+94}(1994).   Velocity data from Chu \markcite{C97}(1997) 
shows a complex expansion structure.  Due to a lack of HRI coverage of this 
object, we have used the \rosat\ PSPC image in this atlas. Its elongated 
X-ray morphology shows an interesting similarity to that of N86. 

{\it N 86}\ (Fig.\,1v). This remnant shows optical extensions to the north 
and south, including filaments that appear to be streaming out from the SNR.
\oiii\ and \sii\ images taken by Lasker \markcite{L77}(1977) show 
northern and southern extensions of N86 that might represent breakouts. 
The elongated X-ray morphology follows these optical features fairly
closely, suggesting that the irregular X-ray structure is due to
breakouts of the hot gas into cavities in the surrounding ISM 
\markcite{W+98}(Paper II).   Due to a lack of HRI coverage of this 
object, we have used the \rosat\ PSPC image in this atlas.

{\it N 186D}\ (No \rosat\ data available).  Due to confusion with the
superposed N 186E region, this SNR has been difficult to observe. A
velocity study by Laval \et\ \markcite{L+89}(1989) was able to 
distinguish the expanding shell of N 186D.  A later interferometric 
study by Rosado \et\ \markcite{R+90}(1990) suggested that that N 186D 
is interacting with the N 186E shell, and that N 186E is an older 
``fossil" SNR now photoionized by internal stars.

{\it DEM L 71}\ (Fig.\,1a). This SNR is Balmer-line dominated at optical
wavelengths \markcite{S+91}(\eg\ Smith \et\ 1991). The X-ray morphology of 
this remnant is interesting for a seeming similarity to the ``barrel shape" 
seen in the radio emission of some Galactic SNRs \markcite{G98}(\eg\ Gaensler 
1998), although it has a bright central patch as well. X-ray spectroscopy 
of this SNR is provided by \markcite{H+98}Hughes \et\ (1998).

{\it N 23}\ (Fig.\,1w).  This object is elongated, with the southeastern limb 
being particularly bright in X-rays.  As the X-ray morphology corresponds
quite well with that in the radio \markcite{DM98}(Dickel \& Milne 1998), it 
seems likely that the morphology is due to an encounter of the shock front 
with higher-density material to the southeast.  X-ray spectroscopy of this 
SNR is provided by \markcite{H+98}Hughes \et\ (1998).

{\it SNR 0509$-$675}\ (Fig.\,1b). This young, Balmer-line dominated SNR shows 
a clear shell structure. It has a probable shock velocity of over 2000 km 
s$^{-1}$ \markcite{S+91}(Smith \et\ 1991), making it one of the youngest of 
the LMC remnants.  A study by Hughes \et\ \markcite{H+95}(1995) suggests that 
this SNR is the result of a Type Ia SN explosion, based on characteristic 
lines in the X-ray spectra.

{\it N 103B}\ (Fig.\,1c). Although Mathewson \et\ \markcite{M+83}(1983) 
classified N103B as having centrally concentrated X-ray emission, an 
examination at higher resolution and sensitivity suggests that the remnant 
is in fact shell-type, and the brighter X-ray region is not central, but in 
fact located near the limb of the SNR.  This offset is easier to see in the 
radio images \markcite{DM95}(Dickel \& Milne 1995), which show a well-defined 
shell brightened to the west.  A study by Hughes \et\ \markcite{H+95}(1995) 
suggests that this SNR, although not Balmer-line dominated at optical
wavelengths, is nonetheless the result of a Type Ia SN explosion, based on 
characteristic lines in the X-ray spectra.

{\it SNR 0513$-$692}\ (No \rosat\ data available). Chu \& Kennicutt 
\markcite{CK88b}(1988b) suggest that this SNR may be the result of
a Type I explosion, on the basis of its nearby environment, which 
includes the \hii\ regions DEM L 108 and DEM L 109, and the OB 
association LH 35.

{\it SNR 0519$-$690}\ (Fig.\,1d). This young, Balmer-line dominated SNR 
has a shell structure with an apparent gap to the southeast. The X-ray 
morphology is similar to that in radio images \markcite{DM95}(Dickel 
\& Milne 1995), although the gap is not as strongly defined in radio.
This SNR has a shock velocity in the range of 1000$-$1900 km s$^{-1}$ 
\markcite{S+91}(Smith \et\ 1991), suggesting it is one of the younger 
LMC remnants.  A study by Hughes \et\ \markcite{H+95}(1995) suggests 
that this SNR is the result of a Type Ia SN explosion, based on 
characteristic lines in the X-ray spectra.

{\it N 120: SNR}\ (Fig.\,1p). This object appears faint at X-ray wavelengths; 
this, combined with its apparently irregular structure, makes it difficult
to classify according to its X-ray morphology.  Due to a lack of HRI coverage 
of this object, we have used the \rosat\ PSPC image in this atlas. Radio image 
for this SNR \markcite{DM98}(Dickel \& Milne 1998) shows a shell-like structure, 
limb-brightened toward the eastern side. This brighter radio region roughly 
corresponds in position to the brightest X-ray emission in the X-ray image, 
casting some doubt on the Centrally Brightened classification.   As the 
brighter limb is in the direction of the bulk of the N120 \hii\ region, it 
seems reasonable to suggest that this limb-brightening in radio is due to 
the higher density material of the \hii\ region. Evidence for such 
interaction is strengthened by the velocity profiles presented by
\markcite{R+93}Rosado \et\ (1993).

{\it SNR 0520$-$694}\ (Fig.\,1l). This object, though faint, shows a fairly 
uniform X-ray distribution with some brighter patches.

{\it SNR 0523$-$679}\ (Fig.\,1m). This SNR was discovered in the N44 \hii\
complex by Chu \et\ \markcite{C+93}(1993).  It shows a somewhat patchy X-ray 
emission distributed over the area of the remnant.  The radio image is 
likewise patchy, but the brighter regions of X-ray emission do not appear 
to correlate with those in the radio images \markcite{DM98}(Dickel \& Milne 
1998).

{\it DEM L 175a}\ (No \rosat\ data available). Chu \& Kennicutt \markcite{CK88b}
(1988b) suggest that this SNR may be the result of a Type I explosion, as it is 
part of the DEM L 175 \hii\ region as well as a CO cloud, and the density 
of nearby OB stars is high.

{\it N 49B}\ (Fig.\,1j). This SNR has fairly uniform X-ray emission over a 
well-defined face, with brighter patches on both the face and limb.  Radio 
images show a faint shell with brighter emission in patches over the face of 
the SNR.  While some of these patches appear correlated with brighter X-ray 
emission, other X-ray bright spots appear in areas that are faint at radio 
wavelengths \markcite{DM98}(Dickel \& Milne 1998).  The same mix of 
correlated and uncorrelated areas between X-ray and optical images was noted
by Vancura \et\ \markcite{V+92} (1992).
X-ray spectroscopy of this SNR is provided by \markcite{H+98} Hughes \et\ (1998).

{\it N 49}\ (Fig.\,1g). N 49 is notable for having had an isolated progenitor 
within a cloudy ISM \markcite{S+85}(Shull \et\ 1985).  The X-ray image shows 
emission over the face of this object, with an extended brightening to the 
south, close to the position of a nearby molecular cloud \markcite{B+97}(Banas 
\et\ 1997).  An unresolved ``X-ray hot spot" coincides with the position of 
the soft gamma-ray repeater (SGR) 0525$-$66  \markcite{M+96}(\eg\ Marsden \et\ 
1996).  The radio morphology is very similar to that seen in X-ray emission, 
but neither radio nor optical images suggest a source corresponding to the 
``X-ray hot spot" \markcite{D+95}\markcite{DM98}(Dickel \et\ 1995; Dickel \& 
Milne 1998). X-ray spectroscopy of this SNR is provided by \markcite{H+98} 
Hughes \et\ (1998).

{\it N 132D}\ (Fig.\,1e). This SNR is notable for having oxygen-rich knots of 
optical emission.  The X-ray morphology is very similar to that seen in the
radio \markcite{DM95}(Dickel \& Milne 1995).  A detailed study of the X-ray 
morphology by \markcite{H87} Hughes (1987) concluded that the structure of 
N 132D in X-rays was caused by the interaction of the SN shock with a 
pre-existing cavity in the ISM, and suggested that the presence of central 
X-ray and optical emission implied that the SN ejecta have not yet interacted 
significantly with circumstellar matter. X-ray spectroscopy of this SNR is 
provided by \markcite{H+98} Hughes \et\ (1998).

{\it DEM L 204}\ (No \rosat\ data available). Chu \& Kennicutt \markcite{CK88b}
(1988b) suggest that this SNR may be the result of a Type II explosion, due
to the paucity of nearby \hii\ regions, CO clouds, or OB associations.

{\it SNR 0528$-$692}\ (Fig.\,2b). The X-ray faintness and lack of clear 
definition for this object make classification difficult.  Chu \& Kennicutt 
\markcite{CK88b}(1988b) suggest it as a result of a type I explosion, as
the SNR is within the \hii\ region DEM L 210 and the density of nearby
OB stars is very high.

{\it N 206}\ (Fig.\,1o). This object is the best candidate for a centrally 
brightened SNR. Chu \& Kennicutt \markcite{CK88a}(1988a) find a continuous 
distribution of velocities, suggesting the presence of material that has 
fallen considerably behind the shock front.  

{\it SNR 0534$-$699}\ (Fig.\,1n). This object, while showing a well-defined 
shell structure in X-ray emission, also shows substantial emission over the
face of the SNR.  

{\it DEM L 238}\ (Fig.\,2a). The total \rosat\ observation time on this 
object was less than 6 kiloseconds. As a result, the emission is difficult 
to detect.  This SNR has been detected at radio wavelengths by Mills \et\ 
\markcite{MT+84}(1984), but has not been extensively studied.

{\it N 63A}\ (Fig.\,1h).  While X-ray and radio images of N 63A show a nearly 
complete shell, the optical emission is confined to three bright ``lobes" 
in a clover-shaped pattern within the X-ray/radio shell. Two of these lobes 
have been shocked by the SNR; the third lobe is a photoionized region 
\markcite{L+95}(Levenson \et\ 1995). These three optical lobes are correlated 
with a local minimum in the X-ray emission, caused by the shadowing effect of 
these lobes \markcite{C97}(Chu 1997). Hubble Space Telescope images of N 63A 
have revealed small shocked cloudlets in the X-ray emission region 
\markcite{C+98}(Chu \et\ 1998).  X-ray spectroscopy of this SNR is provided 
by \markcite{H+98} Hughes \et\ (1998).

{\it DEM L 241}\ (No \rosat\ data available).  A study of an echelle spectrum
of this SNR by Chu \markcite{C97}(1997) shows a relatively small expansion 
velocity and a large intrinsic velocity width, implying that the SNR may
be expanding within a stellar-wind blown cavity.

{\it SN 1987A}\ (Fig.\,1t).  The shock front from this SN is thought to have
recently begun to interact with circumstellar material ejected by its 
progenitor \markcite{S+98}(\eg\ Sonneborn \et\ 1998) and can therefore be 
classified as a very young supernova remnant. The X-ray emission from this 
object is unresolved at the spatial resolution of the HRI.

{\it Honeycomb SNR}\ (Fig.\,1t). This object was identified as a multi-loop 
nebula by \markcite{W92}Wang (1992) and suggested as the result of an SNR 
shock by Meaburn \et\ \markcite{M+93}\markcite{M+95}(1993, 1995).  A 
detailed X-ray and optical study of this object has been carried out by 
Chu \et\ \markcite{C+95}(1995).

{\it DEM L 249}\ (Fig.\,2d). The total \rosat\ observation time on this 
object was less than 6 kiloseconds. As a result, the X-ray emission is 
difficult to detect.  This SNR has been detected at radio wavelengths by 
Mills \et\ \markcite{MT+84}(1984), but has not been extensively studied.

{\it N 157B}\ (Fig.\,1s).  This object is one of the largest known 
Crab-type SNRs \markcite{D+94} (Dickel \et\ 1994).  An interior pulsar has 
recently been identified within this remnant (\markcite{M+98}Marshall \et\ 
1998; \markcite{WG98} Wang \& Gotthelf 1998).  While most of the diffuse 
emission can be attributed to a synchrotron nebula surrounding the central 
pulsar, there appears to be a significant extension away from the point 
source, forming an odd ``head" and ``tail" structure.  While ``tail" could 
simply be part of the pulsar-excited material, it may also be faint emission 
from the outer shell.

{\it N 158A}\ (Fig.\,1r).  This remnant is known to contain a pulsar,
first identified by \markcite{S+84} Seward, Harnden \& Helfand (1984).
While the X-ray emission is dominated by the 
central point source, a clear shell structure is faintly visible in the 
surrounding area. This structure was first noted by Seward \& Harnden 
\markcite{SH94}(1994) in an extensive analysis of the X-ray emission from 
N 158A.  A maximum in the \ha\ emission is correlated with the central point 
source; the shell emission fills an area between two bright \ha\ filaments,
but appears to be unrelated to them (\eg\ \markcite{C+92} Caraveo \et\ 1992).
Recent observations by Caraveo \et\ \markcite{CMB98} (1998) were able to 
resolve the \ha\ emission around this SNR into a ring, with strong 
[\ion{N}{2}] emission.  It is suggested that this ring was formed by the 
pre-supernova phase and later ``swept over" by the expanding SN shock.

{\it SNR 0540$-$697}\ (Fig.\,2e). This object is very close to the X-ray binary 
LMC X-1, and therefore its X-ray emission is largely confused with that 
of the binary.  It was identified as an SNR by its kinematics, soft X-ray
emission, and high \sii /\ha\ ratio \markcite{C+97}(Chu \et\ 1997).

{\it DEM L 299}\ (Fig.\,2c). This object appears to be quite faint at X-ray 
wavelengths.  Even a relatively long (32 ksec) exposure only reveals
a few faint clumps of emission.

{\it DEM L 316 A}\ (Fig.\,1f). Although there appears to be X-ray emission 
over the entire face of this remnant, the emission is dominated by a 
bright X-ray region to the southwest.  Based on the correlation of this 
hotspot with the optical edge of, and radio polarization change in, its 
neighbor DEM L 316 B, Williams \et\ \markcite{W+97}(Paper I) suggest that this 
X-ray bright region is due to a collision between the two SNR shells.

{\it DEM L 316 B}\ (Fig.\,1f). It is notable that the X-ray emission seen 
in DEM L 316 B is well inside the optically bright shell \markcite{W+97}
(Paper I).  This suggests that the X-ray emission from this remnant, as 
opposed to the younger shell-type remnants, is {\it not}\ generated at the 
outer (or reverse) shock, but is instead arising from material within 
the remnant interior.

{\it SNR 0548$-$704}\ (Fig.\,1k). This Balmer-line dominated SNR shows patchy 
X-ray emission incompletely covering the face of the remnant.  Its shock
velocity is relatively rapid, in the 500$-$1100 km s$^{-1}$ \markcite{S+91}
(Smith \et\ 1991) range, which raises the question of why the outer shell 
is not seen more clearly in X-rays.

\clearpage
\begin{figure}
\section{Figure Captions}

\figcaption{\rosat\ images for which the X-ray data were sufficient for 
morphological classification.  All but (q) and (v) were taken with the 
HRI; (q) and (v) used the PSPC.  The scales and image resolutions differ
somewhat for each image; contours are marked on the image, and a black 
spot to the lower right of each image indicates the effective resolution 
(the smoothed half-power beam width). Contours in these images are set as 
multiples of 2 $\sigma$\ above the background, beginning at 2 $\sigma$\ and 
doubling thereafter.  Where X-ray emission is particularly strong, the 
contour levels begin at 20 $\sigma$\ above the background and double 
thereafter.  In cases such as N 158A (r), which contain both bright 
pointlike and faint diffuse emission, both systems are used. See Table 3 
for image statistics.}

\figcaption{\rosat\ HRI images for which the X-ray data were not 
sufficient for morphological classification.  Contours as in Figure 1.}

\figcaption{Cartoon showing the schematic stages of development 
of SNRs in a cloudy medium.  Striped regions are those from which X-ray
emission is expected; while solid regions indicate denser material, from
which optical emission may be observed.  The relative darkness of the 
shading indicates areas of higher (darker) and lower (lighter) density.}

\figcaption{Plot of the X-ray sizes of LMC SNRs. The symbols
represent the different morphological classifications used in this 
paper.}

\figcaption{\ha\ images taken using the 1.5 meter telescope at the 
Cerro Tololo Inter-American Observatory. Images are displayed with a linear
grayscale.  Exceptions are (d) and (f), which are displayed with a 
square-root grayscale to bring out faint emission.}

\figcaption{Radio images taken with the Compact Array of the 
Australia Telescope. Most images are displayed with a linear grayscale;
(c) is displayed with a square-root grayscale to bring out faint emission.}

\end{figure}

\clearpage

\begin{deluxetable}{llccccccc}
\tablewidth{0pc}
\tablecaption{\rosat\ Observations of LMC SNRs}
\tablehead{
\colhead{Object}  &     
\colhead{Name}      &
\colhead{R. A.}      & 
\colhead{Dec.}      & 
\colhead{Inst} &
\colhead{Seq} &
\colhead{ExpT} & 
\colhead{Fig} \\
\colhead{}      & 
\colhead{(in lit.)\tablenotemark{a}}      &
\colhead{(J2000)}      & 
\colhead{(J2000)}      & 
\colhead{} &
\colhead{No} &
\colhead{(ks)} & 
\colhead{No} & 
}

\startdata
0453$-$685 & SNR 0453$-$685 & 04 53 37 & -68 29.5 & HRI & 500231 & 28.6 & 1i \nl
0453$-$669 & SNR in N 4 & 04 53 14 & -66 55.2\tablenotemark{b} & PSPC & 500263  & 12.7 & 1x \nl
0454$-$665 & N 11L & 04 54 48 & -66 25.6 & HRI & 900321 & 32.2 & 1u \nl
0454$-$672 & SNR in N 9 & 04 54 33 & -67 12.8\tablenotemark{b} & PSPC & 500263  & 12.7 & 1q \nl
0455$-$687 & N 86 & 04 55 44 & -68 39.0  & PSPC & 500258 & 12.7 & 1v \nl
0505$-$679 & DEM L 71 & 05 05 42 & -67 52.7  & HRI & 500228 & 10.1 & 1a \nl
0506$-$680 & N 23 & 05 05 56 & -68 01.8  & HRI & 500229 & 10.3 & 1w \nl
0509$-$675 & SNR 0509$-$675 & 05 09 31 & -67 31.3  & HRI & 500177 & 2.8 & 1b \nl
0509$-$687 & N 103B & 05 09 00 & -68 43.6 & HRI & 500174 &  7.1 & 1c \nl
0519$-$690 & SNR 0519$-$690 & 05 19 35 & -69 02.1 & HRI & 500171 & 11.3 & 1d \nl
0519$-$697 & N 120 & 05 18 42 & -69 39.5 & PSPC & 180033 & 16.0  & 1p \nl
0520$-$694 & SNR 0520$-$694 & 05 19 46 & -69 26.0 & HRI & 600919 & 18.1 & 1l \nl
0523$-$679 & SNR in N44 & 05 23 18 & -67 56.0\tablenotemark{b} & HRI & 600913 & 109. & 1m \nl
0525$-$660 & N 49B & 05 25 26 & -65 59.3 & HRI & 500172 & 13.3 & 1j \nl
0525$-$661 & N 49 & 05 26 00 & -66 05.0& HRI & 500172 & 13.3 & 1g \nl
0525$-$696 & N 132D & 05 25 02 & -69 38.6 & HRI & 500002 & 26.4 & 1e \nl
0528$-$692 &  SNR 0528$-$692  & 05 27 39 & -69 12.0 & HRI & 600641 & 26.5 & 2b \nl
0532$-$710 & N 206 & 05 31 57 & -71 00.2 & HRI & 600781 & 49.2 & 1o \nl
0534$-$699 & SNR 0534$-$699 & 05 34 00 & -69 54.9 & HRI & 600780 & 14.3 & 1n \nl
0534$-$705 & DEM L 238 & 05 34 15 & -70 33.7 & HRI & 400352 &  5.5 & 2a \nl
0535$-$660 & N 63A & 05 35 44 & -66 02.1 & HRI & 500173 &  2.7 & 1h \nl
SN 1987A & SN 1987A & 05 35 28 & -69 16.2 & HRI & 500407 & 39.0 & 1t \nl
0536$-$693 & Honeycomb & 05 35 44 & -69 18.2 & HRI & 500407 & 39.0 & 1t \nl
0536$-$706 & DEM L 249 & 05 36 06 & -70 38.7 & HRI & 400352 &  5.5 & 2d \nl
0538$-$691 & N 157B & 05 37 48 & -69 10.3 & HRI & 500036 &  8.8 & 1s \nl
0540$-$693 & N 158A & 05 40 11 & -69 20.0 & HRI & 150008 & 18.1 & 1r \nl
0540$-$697 & SNR in N 159 & 05 39 58 & -69 44.0\tablenotemark{b} & HRI & 600779 & 26.2 & 2e  \nl
0543$-$689 & DEM L 299 & 05 43 15 & -68 59.4 & HRI & 500235 & 31.9 & 2c \nl
0547$-$697 & DEM L 316 A & 05 47 20 & -69 41.3 & HRI & 500232 & 17.7 & 1f \nl
0547$-$697 & DEM L 316 B & 05 47 02 & -69 43.0  & HRI & 500232 & 17.7 & 1f \nl
0548$-$704 & SNR 0548$-$704 & 05 47 48 & -70 24.8 & HRI & 500233 & 25.1 & 1k \nl
\enddata

\tablenotetext{a}{We include the designations of these objects most 
commonly found in the literature.  Note that many of these designations
are technically incorrect, as the SNRs are referred to by names that
actually refer to nearby or surrounding \hii\ regions (\eg\ N 206).}

\tablenotetext{b}{Coordinates cited from  Chu \et\ 1993 (in N44);
Smith \et\ 1994 (in N4, N9); Chu \et\ 1997 (in N159).}

\end{deluxetable}

\begin{deluxetable}{llccccc}
\tablewidth{0pc}
\tablecaption{X-ray Brightnesses and Sizes of LMC SNRs}
\tablehead{
\colhead{Object}  &     
\colhead{Name}      &
\colhead{Count Rate} &
\colhead{Avg. Bright} & 
\colhead{X-ray} \\
\colhead{}      & 
\colhead{}      & 
\colhead{(ct/s)} &
\colhead{(ct/s/$\Box$\amin)} &
\colhead{Size} &
}

\startdata
0453$-$685 & SNR 0453$-$685 & 0.20 $\pm$ 0.005 & 0.046 & 2\ampt4$\times$2\ampt3 \nl
0453$-$669 & SNR in N 4 & 0.028 $\pm$ 0.002\tablenotemark{a} & 0.0020\tablenotemark{a} & 5\ampt3$\times$3\ampt3 \nl
0454$-$665 & N 11L & 0.0096 $\pm$ 0.002 & 0.0040 & 1\ampt9$\times$1\ampt6  \nl
0454$-$672 & SNR in N 9 & 0.059 $\pm$ 0.003\tablenotemark{a} & 0.0059\tablenotemark{a} & 4\ampt9$\times$2\ampt6 \nl
0455$-$687 & N 86 & 0.053 $\pm$ 0.003\tablenotemark{a} & 0.0040\tablenotemark{a} & 6\ampt1$\times$2\ampt8 \nl
0505$-$679 & DEM L 71 & 0.65 $\pm$ 0.009 & 0.57 & 1\ampt2$\times$1\ampt2 \nl
0506$-$680 & N 23 & 0.41 $\pm$ 0.007 & 0.25 & 1\ampt55$\times$1\ampt4  \nl
0509$-$675 & SNR 0509$-$675 & 0.20 $\pm$ 0.009 & 0.71 & 0\ampt6$\times$0\ampt6  \nl
0509$-$687 & N 103B & 0.72 $\pm$ 0.01 & 2.0 & 0\ampt65$\times$0\ampt7  \nl
0519$-$690 & SNR 0519$-$690 & 0.60 $\pm$ 0.008 & 1.4 & 0\ampt75$\times$0\ampt75 \nl
0519$-$697 & N 120 & 0.025 $\pm$ 0.003\tablenotemark{a} & 0.0027\tablenotemark{a} &3\ampt0$\times$4\ampt0  \nl
0520$-$694 & SNR 0520$-$694 & 0.044 $\pm$ 0.003 & 0.0098 & 2\ampt6$\times$2\ampt2  \nl
0523$-$679 & SNR in N44 & 0.013 $\pm$ 0.001 & 0.00096 & 4\ampt3$\times$4\ampt0  \nl 
0525$-$660 & N 49B & 0.57 $\pm$ 0.007 & 0.11 & 2\ampt8$\times$2\ampt4  \nl
0525$-$661 & N 49 & 0.87 $\pm$ 0.008 & 0.57  & 1\ampt4$\times$1\ampt4   \nl
0525$-$696 & N 132D & 4.5 $\pm$ 0.01 & 1.5 & 2\ampt2$\times$1\ampt7 \nl
0528$-$692 & SNR 0528$-$692 & 0.0055 $\pm$ 0.001 & $<$ 0.0011 & $\sim$2\ampt5 \nl
0532$-$710 & N 206 & 0.039 $\pm$ 0.01 & 0.0038 & 3\ampt7$\times$3\ampt5  \nl
0534$-$699 & SNR 0534$-$699 & 0.086 $\pm$ 0.004 & 0.020 & 2\ampt4$\times$2\ampt3 \nl
0534$-$705 & DEM L 238 & 0.018 $\pm$ 0.003 & $<$ 0.0037  & $\sim$2\ampt5  \nl
0535$-$660 & N 63A & 2.8 $\pm$ 0.03 & 2.0 & 1\ampt35$\times$1\ampt35 \nl 
SN 1987A   & SN 1987A & 0.0013 $\pm$ 0.0003 & $>$ 0.16 & $<$ 0\ampt1 \nl 
0536$-$693 & Honeycomb & 0.0057 $\pm$ 0.0009 & 0.0061 & 1\ampt7$\times$0\ampt7   \nl
0536$-$706 & DEM L 249 & 0.014 $\pm$ 0.003 & 0.0010  & 5\ampt2$\times$3\ampt3  \nl
0538$-$691 & N 157B & 0.061 $\pm$ 0.003 & 0.081  & 1\ampt2$\times$0\ampt8   \nl
0540$-$693 & N 158A & 0.25 $\pm$ 0.004 & 0.18  & 1\ampt5$\times$1\ampt2  \nl
0540$-$697 & SNR in N 159 & 0.020 $\pm$ 0.001\tablenotemark{b} & $<$ 0.025\tablenotemark{b} &  $\sim$1\amin \nl
0543$-$689 & DEM L 299 & 0.046 $\pm$ 0.004 & $<$ 0.0037 & $\sim$4\amin \nl
0547$-$697 & DEM L 316 A & 0.021 $\pm$ 0.002 & 0.0093 & 1\ampt7$\times$1\ampt7  \nl
0547$-$697 & DEM L 316 B & 0.023 $\pm$ 0.002 & 0.0038 & 2\ampt1$\times$3\ampt7  \nl
0548$-$704 & SNR 0548$-$704 & 0.059 $\pm$ 0.002 & 0.022 & 1\ampt9$\times$1\ampt8  \nl
\enddata

\tablenotetext{a}{These SNRs were observed with the \rosat PSPC, 
which has about three times the sensitivity of the HRI to diffuse emission. 
The Count Rate (ct/s) and Average Brightness (ct/s/arcmin$^2$) for each
should be adjusted accordingly.}

\tablenotetext{b}{Confusion exists between the emission
of this SNR and that of LMC X-1.}

\end{deluxetable}

\begin{deluxetable}{llccclccc}
\tablewidth{0pc}
\tablecaption{\rosat\ LMC SNR Image Data}
\tablehead{
\colhead{Object}  &     
\colhead{Name}      &
\colhead{Pix} &
\colhead{Img} &
\colhead{Low Cntr} &
\colhead{$\sigma$} &
\colhead{Contour}  \\
\colhead{} &
\colhead{} &
\colhead{Size} &
\colhead{Res\tablenotemark{a}} &
\colhead{(ct/pix)}&
\colhead{(ct/pix)}&
\colhead{Levels ($\sigma$)} 
}

\startdata
0453$-$685 & SNR 0453$-$685 & 2\asec & 11\asec & 0.27 & 0.060 & 2, 4, 8, 16, 32, 64 \nl
0453$-$669 & SNR in N 4 & 5\asec & 38\asec & 0.16 & 0.030 & 2, 4, 8 \nl
0454$-$665 & N 11L & 5\asec & 24\asec & 1.3 & 0.15 & 2, 4, 8, 16 \nl
0454$-$672 & SNR in N 9 & 5\asec & 38\asec & 0.18 & 0.045 & 2, 4, 8, 16, 32 \nl
0455$-$687 & N 86 & 5\asec & 38\asec & 0.14 & 0.039 & 2, 4, 8, 16 \nl
0505$-$679 & DEM L 71 & 1\asec & 7\asec & 0.48 & 0.023 & 20, 40, 80 \nl
0506$-$680 & N 23 & 1\asec & 7\asec  & 0.074 & 0.024 & 2, 4, 8, 16, 32, 64 \nl
0509$-$675 & SNR 0509$-$675 & 1\asec & 7\asec  & 0.44 & 0.0055 & 20, 40, 80, 160 \nl
0509$-$687 & N 103B & 1\asec & 7\asec  & 0.38 & 0.018 & 20,40,80,160,320,640 \nl
0519$-$690 & SNR 0519$-$690 & 1\asec & 7\asec & 0.47 & 0.022 & 20, 40, 80, 160, 320 \nl
0519$-$697 & N 120 & 5\asec & 38\asec & 0.48 & 0.082 & 2, 4, 8 \nl
0520$-$694 & SNR 0520$-$694 & 2\aspt5 & 13\asec  & 0.34 & 0.076 & 2, 4, 8 \nl
0523$-$679 & SNR in N44 & 5\asec & 24\asec  &  4.3 & 0.26 & 2, 4, 8  \nl 
0525$-$660 & N 49B  & 1\asec & 7\asec & 0.060 & 0.019 & 2, 4, 8, 16, 32 \nl
0525$-$661 & N 49 & 1\asec & 7\asec & 0.37 & 0.14 & 2, 8, 16, 32 \nl
0525$-$696 & N 132D & 1\asec & 7\asec & 2.4 & 0.11 & 20, 40, 80, 160 \nl
0528$-$692 & SNR 0528$-$692 & 5\asec & 24\asec  & 1.4 & 0.18 & 2, 4, 8 \nl
0532$-$710 & N 206 & 5\asec & 24\asec & 1.9 & 0.17 & 2, 4, 8, 16 \nl
0534$-$699 & SNR 0534$-$699 & 2\aspt5 & 13\asec & 0.22 & 0.048 & 2, 4, 8, 16 \nl
0534$-$705 & DEM L 238 & 5\asec & 24\asec & 0.28 & 0.059 & 2, 4, 8 \nl
0535$-$660 & N 63A & 1\asec & 7\asec & 0.27 & 0.21 & 2, 8, 16, 32  \nl 
SN 1987A & SN 1987A & 1\asec & 7\asec & 0.37 & 0.079 & 2, 4, 8, 16  \nl 
0536$-$693 & Honeycomb & 1\asec & 7\asec & 0.37 & 0.079 & 2, 4, 8, 16 \nl
0536$-$706 & DEM L 249 & 5\asec & 24\asec & 0.24 & 0.052 & 2, 4, 8 \nl
0538$-$691 & N 157B & 1\asec & 7\asec & 0.052 & 0.017 & 2, 4, 8, 16, 32  \nl
0540$-$693 & N 158A & 1\asec & 7\asec & 0.093 & 0.029 & 2,4,8,20,40,... 640 \nl
0540$-$697 & SNR in N 159 & 5\asec & 24\asec  & 1.7 & 0.22 & 2, 4, 8, 16, 32 \nl
0543$-$689 & DEM L 299 & 5\asec & 24\asec & 1.6 & 0.19 & 2, 4 \nl
0547$-$697A & DEM L 316 A & 2\aspt5 & 13\asec  & 0.21 & 0.048 & 2, 4, 8, 16  \nl
0547$-$697B & DEM L 316 B & 2\aspt5 & 13\asec  & 0.21 & 0.048 & 2, 4, 8, 16  \nl
0548$-$704 & SNR 0548$-$704 & 1\asec & 7\asec & 0.085 & 0.025 & 2, 4, 8, 16 \nl
\enddata

\tablenotetext{a}{All images are also smoothed with a Gaussian of
$\sigma$=2 pixels.}

\end{deluxetable}

\begin{deluxetable}{llll}
\tablewidth{0pc}
\tablecaption{Classification of LMC SNRs}
\tablehead{
\colhead{Object}      & 
\colhead{Name}      &
\colhead{Category} &
\colhead{Features} 
}

\startdata
0450$-$709 & SNR 0450$-$709 & No data\tablenotemark{a} & -- \nl 
0453$-$685 & SNR 0453$-$685 & Diffuse Face &  \nl
0453$-$669 & SNR 0453$-$669 & Irregular & Bright Limb \nl
0454$-$665 & N 11L & Irregular & Elongated Feature \nl
0454$-$672 & SNR 0454$-$672 & Centrally Brightened? & Elongated Feature \nl
0455$-$687 & N 86 & Irregular & Elongated Features \nl
0500$-$702 & N 186D & No data  & -- \nl
0505$-$679 & DEM L 71 & Shell & Gap, Non-Limb Feature \nl
0506$-$680 & N 23 & Irregular & Bright Limb, Non-Limb Feature \nl
0509$-$675 & SNR 0509$-$675 & Shell & Limb Hotspot \nl
0509$-$687 & N 103B & Shell & Limb Hotspot \nl
0513$-$692 & SNR 0513$-$692  & No data & --  \nl
0519$-$690 & SNR 0519$-$690 & Shell  & Gap \nl
0519$-$697 & N 120 & Centrally Brightened? & --  \nl
0520$-$694 & SNR 0520$-$694 & Diffuse Face &  \nl
0523$-$679 & SNR 0523$-$679 & Diffuse Face & Elongated Feature  \nl
0524$-$664 & DEM L 175a & No data  & --  \nl
0525$-$660 & N 49B & Diffuse Face & Non-Limb Feature \nl
0525$-$661 & N 49 & Diffuse Face & Non-Limb Feature \nl
0525$-$696 & N 132D & Shell & Non-Limb Feature  \nl
0527$-$658 & DEM L 204 & No data & --  \nl
0528$-$692 & SNR 0528$-$692 & Unclassified\tablenotemark{b}  & --   \nl
0532$-$710 & N 206 & Centrally Brightened &  \nl
0534$-$699 & SNR 0534$-$699 & Diffuse Face  &  \nl
0534$-$705 & DEM L 238 & Unclassified &  \nl
0535$-$660 & N 63A & Diffuse Face & Limb Hotspot \nl
0536$-$676 & DEM L 241 & No data  & -- \nl
SN 1987A & SN 1987A & Special\tablenotemark{c}  & Unresolved \nl
0536$-$693 & Honeycomb & Irregular &  \nl
0536$-$706 & DEM L 249 & Unclassified  & --   \nl
0538$-$691 & N 157B & Peaked Emission & Elongated Feature \nl
0540$-$693 & N 158A & Peaked Emission & Unclassified shell \nl
0540$-$697 & SNR 0540$-$697 & Unclassified & Obscured by LMC X-1  \nl
0543$-$689 & DEM L 299 & Unclassified  & --  \nl
0547$-$697A & DEM L 316 A & Diffuse Face & Bright Limb/Hotspot \nl
0547$-$697B & DEM L 316 B & Shell & Elongated Feature \nl
0548$-$704 & SNR 0548$-$704 & Diffuse Face & Gap? \nl
\enddata

\tablenotetext{a}{No data: These objects have not been observed, or
   were not detected, with the \rosat\ instruments.}

\tablenotetext{b}{Unclassified: These objects do not show an 
   unambiguous detection in the \rosat\ observations.} 

\tablenotetext{c}{SN 1987A is still changing on yearly 
timescales, and is not resolved by the \rosat\ HRI.}

\end{deluxetable}

\end{document}